\documentclass[prl,amsmath,twocolumn,showpacs]{revtex4-1}

\usepackage{epsfig}
\usepackage{amsfonts} 
\usepackage{nicefrac}
\usepackage{color} 
\usepackage{hyperref}
\setcitestyle{square}
\usepackage{subfigure} 
\usepackage{natbib} 
\usepackage{graphicx,amssymb} 
\usepackage{bm}

\newcommand{\ignore}[1]{} 

\def\eq#1{${#1}$}

\usepackage{gensymb} 
\setcitestyle{square}
\usepackage{subfigure}

\usepackage{enumitem}

\begin{document}

\title{Reply to 
Comment on \\
 ``Large fluctuations for spatial diffusion of cold atoms''} 
\author{Erez Aghion\footnote{Correspondence to Erez Aghion: ErezAgh5@gmail.com}, David A. Kessler, Eli Barkai}
\affiliation{ Department of Physics, Institute of Nanotechnology and Advanced Materials, Bar-Ilan University, Ramat-Gan 52900, Israel}

%%%%%%%%%%%%%%%%%%%%%%%%%%%%%%%%%%%%%%%%%%%%%%%%%%%%%%%%%%%%%%%%%%%%%%%%%%%%%%%
%
% A B S T R A C T 
%
%%%%%%%%%%%%%%%%%%%%%%%%%%%%%%%%%%%%%%%%%%%%%%%%%%%%%%%%%%%%%%%%%%%%%%%%%%%%%%%
\begin{abstract}
 We provide a reply to a comment by I. Goychuk  arXiv:1708.04155, version v2 from $27$ Aug $2017$ (not under active consideration with Phys. Rev. Lett.) \cite{goychuk2017comment}, on our Letter E. Aghion, D. A. Kessler and E. Barkai, \textbf{Phys. Rev. Lett., 118}, 260601 (2017) \cite{aghion2017large}.  
\end{abstract}

%%%%%%%%%%%%%%%%%%%%%%%%%%%%%%%%%%%%%%%%

\iffalse 
\begin{center}
Reply to 
Comment on \\
 ``Large fluctuations for spatial diffusion of cold atoms''
\end{center}

\textit{ We provide a reply to a comment by I. Goychuk  arXiv:1708.04155v2, version v2 from $27$ Aug $2017$ (not under active consideration with PRL), on our Letter E. Aghion, D. A. Kessler and E. Barkai, \textbf{PRL, 118}, 260601 (2017) \cite{aghion2017large}. }

\vspace{1cm} \fi

\maketitle{}

{\em Background:} Our work on the spatial diffusion of cold atoms contains two
parts. At first we investigated the typical fluctuations,
giving precise predictions on the bulk spreading of the
 packet of particles \cite{kessler2012theory,barkai2014area},
using a semiclassical formalism.
Among the predictions of the model is the L\'evy spreading of the atoms
in a shallow optical lattice, first suggested in  1996 by Marksteiner et al. \cite{marksteiner1996anomalous}.
This behavior was observed
by Sagi et al. 
\iffalse[Sagi et al. \textbf{PRL}, 2012]\fi\cite{sagi2012observation},
in an experiment which for the first time clearly
demonstrated the non-Gaussian
dynamics of the cloud of atoms. 
 More recently
we have investigated theoretically the  far tails of the particles density. While there exists available experimental data on
the typical bulk spreading of the atoms, currently the experimental results
cannot say much on the far tails. The questions raised in the
comment, while not directly related to the topic of our most recent publication \cite{aghion2017large},
are certainly relevant for our previous works. The reader who
wishes to explore this issue, should  first consult our
previous papers on the diffusion of Sisyphus-cooled atoms to gain a well informed opinion. 
We now turn to provide a reply to the comment.

 In the experiment whose results are cited in the comment in an attempt to discredit the model under study, 
the total number of particles is not conserved. Our theory deals with
a normalized density where the total number
of particles is fixed. In the experiment  only 10\%-30\% of the particles were
preserved due to radial losses. To quote from \cite{sagi2012observation}: \textit{``Nevertheless, we use the data only as long as at least \eq{10}\% of the atoms remain in the trap. Increasing this number to \eq{30\%} changes only the $3$ largest points in Fig. 3 by up to \eq{25}\%''}. We believe that a \eq{25}\% change is large enough to make it extremely hard for quantitative comparison between theory and experiment (of course the effect could have been much larger if all $100\%$ of the particles were taken into consideration). The very fast particles in the system 
are possibly 
those particles that are evaporated from the experimental setup. In any case,  it is premature to disqualify the model
based on a single experiment.  The semiclassical description of laser cooled atoms 
 under study and the special friction
force of Sisyphus cooling
 is successful in many of its predictions, e.g. the velocity distribution
and the transition from Gaussian to L\'evy dynamics of the spatial 
spreading. Read more for example  in \iffalse[Cohen-Tanudji and Philip, \textbf{Phys. Today} (1990); Lutz, \textbf{PRA} (2003); Lutz and Renzoni, \textbf{Nature Phys.} (2013); Sagi et al., \textbf{PRL} (2012)]\fi\cite{sagi2012observation,cohen1990new,lutz2003anomalous,lutz2013beyond}. The whole comment is misdirected, in the sense that
we use a well established model\iffalse (from the mid 1990s)\fi,
 and make predictions based on it,  
with the hope that a future experiment will clarify its validity by careful
data analysis.

  The author of the comment also wonders with respect to the meaning
of the word cold in the field of cold-atom physics. 
Again the discussion is misdirected. The literature is full
of deliberations on this issue.  The author used the work of two of us on the
variance of the momentum of the particles \cite{kessler2010infinite} to
manufacture a problem. 
Very briefly, our theory \cite{aghion2017large}, based on the semiclassical description, is
valid for  $D<1$. This includes, but is  not limited to the
case where the variance of the momentum is increasing with time, as
we have demonstrated  a while ago in \iffalse [Kessler and Barkai, \textbf{PRL} (2010)]\fi\cite{kessler2010infinite}.
More generally, let  us
 quote from Cohen-Tannoudji's Nobel Prize lectures: \textit{``It is usual in laser cooling to define an effective temperature
$T$ in terms of the half width $\delta p$
(at \eq{1/\sqrt{e}}) of the momentum distribution''.}
Thus laser cooled atoms are cold in the sense of a small full width at {$(1/\sqrt{e})$} \iffalse \sout{half}\fi maximum
of the velocity distribution compared with other cooling methods.
 The variance of the momentum might be relatively large, as is well known.
However, the variance
is a measure of the tails of the velocity probability density function.
This issue is well understood and it is certainly not related to our work.
For further explanation of this point, see
the book on L\'evy statistics
and sub-recoil laser cooling by Bardou, Bouchaud, Aspect and Cohen-Tannoudji \cite{bardou2002levy},
which shows how power law statistics and laser cooled atoms live
together in perfect harmony.

 In our Letter we neither use the word temperature nor the symbol $T$.
The atoms are driven by the external laser fields,
and while they can reach slow velocities, we do not make the claim
that these systems are thermal. We emphasized this point
extensively in \iffalse [Dechant et al., \textbf{PRL} (2015)]\fi 
\cite{dechant2015deviations,dechant2016heavy}.
In other words the symbol $T$, usually assigned to temperature and as used
by Goychuk, is  misleading since it usually refers to 
classical systems in thermal equilibrium. 

The author of the comment is also bothered by the fact that our theory predicts a different scaling for the full width at half maximum than for the variance of the spatial distribution (as well as the momentum distribution, needless to say). However, this is not just a prediction of our theory. This is an experimental fact. The fact that Sagi et al. find a L\'evy distribution, whose variance is either infinite, or  totally determined by some cutoff at large scales, perforce 
implies 
 that the two measures of the width have different scaling behaviors. The fact that the author of the comment has not assimilated this simple fact is symptomatic of his failure to understand the inherently anomalous behavior of the system. 
 
  In Fig. $1$ of the comment the diffusion exponent $\delta$, as a function
of the depth of the optical lattice is plotted. 	It is  defined by Goychuk according to
$\langle x^2 (t) \rangle \sim t^{2/\delta}$. In that figure one can also observe the experimental results.
However note that in experiments $\delta$ was not recorded, and the mean-square displacement was not reported.
This is most likely due to high noise to signal ratio in the tails of the packet and the reasons
just discussed in the previous section. The experiment reports the exponent describing the full width at half maximum, hence it follows
that the whole discussion on the value of $c$ in the comment is misleading \footnote{The value of $c=22$ used in our recent letter was taken from \cite{moi1991light,lutz2004power,douglas2006tunable}, and was calculated for a specific two-state model of a Cesium atom. In \cite{sagi2012observation} Sagi et al.  used \eq{{}^{87}}\eq{Rb} atoms, which is also an important distinction, since, as we mentioned in the supplemental material of our letter \cite{aghion2017large}, \eq{c} depends on the transition rates between the particles' energy states. \iffalse Another value, \eq{c= 12.3}, was found for atoms in molasses with a \eq{J_g = 1/2\rightarrow J_e = 3/2} Zeeman transition from the ground state \eq{J_g} to the excited \eq{J_e} in a lin\eq{\perp}lin laser configuration \cite{marksteiner1996anomalous}.\fi See  further discussion for example in \cite{dechant2016heavy}.}

 To conclude let us summarize the way we see the comparison between
theory and experiment.
 The model under study gives many predictions tested in experiments,
e.g. the velocity distribution, see for example \iffalse[Douglas et al., \textbf{PRL} (2006)]\fi\cite{lutz2004power,douglas2006tunable}. 
As we discussed also in \iffalse[Barkai et al., \textbf{PRX} (2014)], Ref.\fi \cite{barkai2014area}, 
the comparison between the theory and experiment in the case of the spatial diffusion of the atoms is only qualitative in the sense that
the model predicts correctly
the transition from Gaussian to L\'evy behavior, but does not 
describe it quantitatively. In fact one of the main reasons of
us exploring in depth the semiclassical description in the current
work, was to make a prediction for the far tails of the density (i.e.
the L\'evy density must be cut off due to the physical reason which is the finite
speed of the particles). 
In  the PRX \cite{barkai2014area} 
we also speculated on the reasons for these observed deviations of experiment from theory.
 One issue, as mentioned, is the
loss of particles in experiments, the second is slightly more subtle.
In the absence of the optical lattice the spreading should be ballistic.
We consider the limit of finite depth of the optical lattice, at long times,
where for shallow lattices the model predicts super-ballistic behavior.
The limits $U_0 \to 0$ and $t \to \infty$, however, do not commute.
In the experiment, the measurement time was fixed and $U_0$ was taken to be small.
Of course it is possible that other reasons are behind the deviations,
like the need to go beyond the semiclassical limit.
All in all, further experiments are needed to check the validity of the
semiclassical theory for shallow lattices. Numerical results
based on more sophisticated models, which take into consideration the lattice structure,
 e.g. \iffalse [Marksteiner et al., \textbf{PRA} (1996); Holtz et al., \textbf{EPL} (2015)]\fi\cite{marksteiner1996anomalous,holz2015infinite}, 
give credence to the semiclassical formalism. 
\\

\textbf{Acknowledgement:} this work was supported by the
Israel Science Foundation.

\bibliographystyle{aipnum4-1}
\bibliography{./bibliography2} 

\end{document}